\documentclass[conference]{IEEEtran}
\IEEEoverridecommandlockouts
\usepackage{cite}
\usepackage{amsmath,amssymb,amsfonts}
\usepackage{algorithmic}
\usepackage{graphicx}
\usepackage{textcomp}
\usepackage{xcolor}
\usepackage[a4paper, total={184mm,239mm}]{geometry}
\usepackage{dblfloatfix}
\usepackage{float}      
\usepackage{capt-of}       
\usepackage{placeins}      
\usepackage{soul}
\usepackage{wrapfig}   

\def\BibTeX{{\rm B\kern-.05em{\sc i\kern-.025em b}\kern-.08em
    T\kern-.1667em\lower.7ex\hbox{E}\kern-.125emX}}
\begin{document}

\title{HALO: Memory-Centric \underline{H}eterogeneous \underline{A}ccelerator with 2.5D Integration for \underline{Lo}w-Batch LLM Inference}

\newcommand{\papername}{HALO }
\newcommand{\papernamewospace}{HALO}

\author{Shubham Negi and Kaushik Roy\\
Elmore Family School of Electrical and Computer Engineering, Purdue University\\
West Lafayette, IN 47907, USA\\
{\tt\small snegi@purdue.edu}
}

\maketitle

\begin{abstract}


The rapid adoption of Large Language Models (LLMs) has driven a growing demand for efficient inference, particularly in latency-sensitive applications such as chatbots and personalized assistants. Unlike traditional deep neural networks, LLM inference proceeds in two distinct phases: the \textit{prefill} phase, which processes the full input sequence in parallel, and the \textit{decode} phase, which generates tokens sequentially. These phases exhibit highly diverse compute and memory requirements, which makes accelerator design particularly challenging. Prior works have primarily been optimized for high-batch inference or evaluated only short input context lengths, leaving the low-batch and long-context regime, which is critical for interactive applications, largely underexplored.

In this work, we propose \papernamewospace, a heterogeneous memory-centric accelerator specifically designed to address the unique challenges of prefill and decode phases in low-batch LLM inference. \papername integrates HBM based Compute-in-DRAM (CiD) with an on-chip analog Compute-in-Memory (CiM), co-packaged using 2.5D integration. To further improve the hardware utilization, we introduce a phase-aware mapping strategy that adapts to the distinct demands of the prefill and decode phases. Compute-bound operations in the prefill phase are mapped to CiM to exploit its high throughput matrix multiplication capability, while memory-bound operations in the decode phase are executed on CiD to benefit from reduced data movement within DRAM. Additionally, we present an analysis of the performance trade-offs of LLMs under two architectural extremes: a fully CiD and a fully on-chip analog CiM design to highlight the need for a heterogeneous design. We evaluate \papername on LLaMA-2 7B and Qwen3 8B models. Our experimental results show that LLMs mapped to \papername achieve up to 18$\times$ geometric mean speedup over AttAcc, an attention-optimized mapping and 2.5$\times$ over CENT, a fully CiD based mapping.

\end{abstract}


\section{Introduction}

\renewcommand{\thefootnote}{}
\renewcommand{\thefootnote}{\arabic{footnote}}


Large Language Models (LLMs) have shown tremendous progress in diverse application types. These include not only tasks in Natural Language Processing (NLP) such as chatbots \cite{chatbot}, text summarization \cite{summarization} and code generation \cite{github2025copilot}, but also image and video processing tasks \cite{imagegen, video}. Beyond generative use cases, LLMs are also increasingly employed in prefill heavy discriminative tasks such as recommendation systems \cite{wang2023enhancing, wu2024survey}, credit verification \cite{son2023beyond}, and data labeling \cite{he2023annollm, lan2024depression}. The inference process of LLMs typically consists of two distinct phases: \textit{prefill} and \textit{decode}. In the prefill phase, the model processes the entire input sequence of length $L_{in}$ (input context length), while in the decode phase it autoregressively generates one token at a time until it produces an output sequence of length $L_{out}$ (output context length). 


\begin{figure}[t]
\centering
\includegraphics[width=0.45\textwidth]{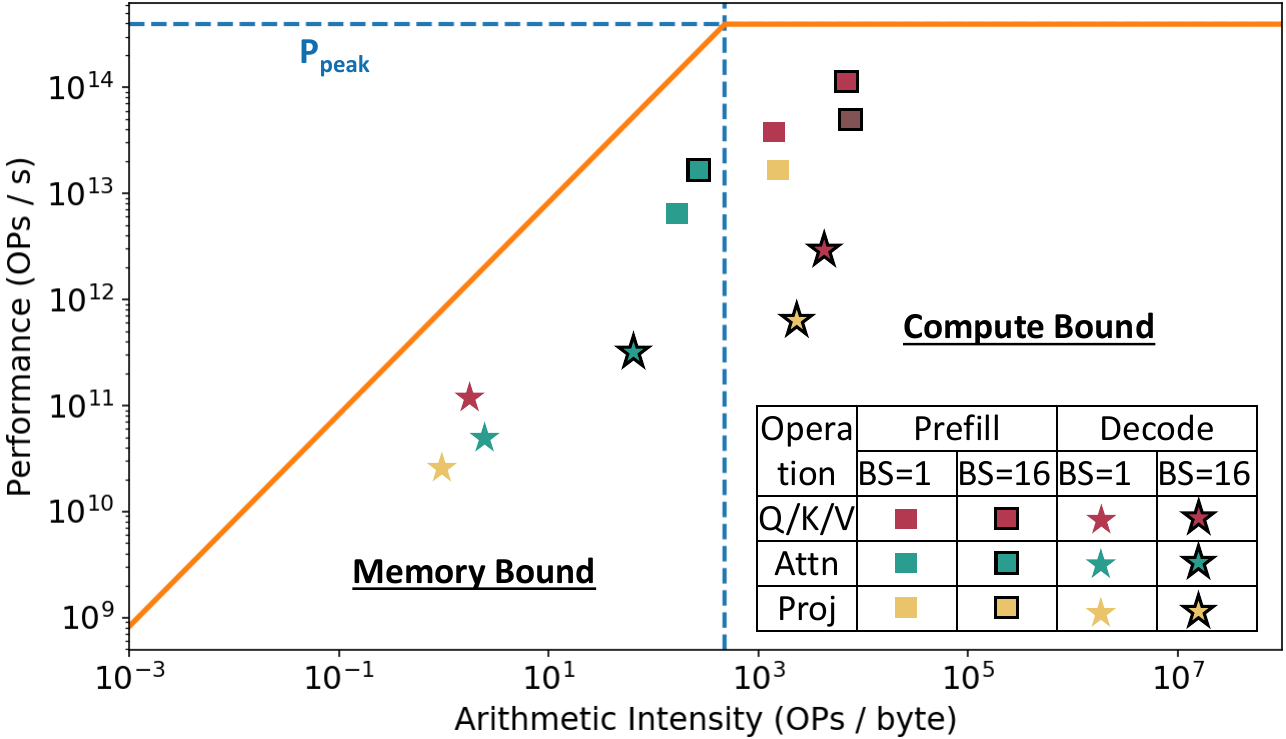}
    \caption{Roofline plot of the CiM accelerator (Table~\ref{spec}) with general matrix-matrix multiplication (GEMM) operations ($L_{in}$=512) from the LLaMA-2 7B model mapped during prefill and decode phases for batch size (BS) 1 and 16, respectively. Prefill GEMMs generally achieve higher arithmetic intensity and approach the compute bound region, while decode GEMMs, especially batch size 1, are memory bound and limited by bandwidth.}

\centering
\label{motivation}
\vskip -0.2 in
\end{figure}

Despite their strong performance, LLMs face significant challenges due to their high memory and compute requirements \cite{chavan2024fasterlighterllmssurvey}. First, the model size itself is large, and the effective memory footprint further grows with the output context length \cite{wu2025inferencescalinglawsempirical}. To address this high memory footprint of LLMs, prior works have explored compression based techniques such as quantization, pruning and neural architecture search \cite{quant1, quant2, pruning, dong2024lpzero, bercovich2025puzzledistillationbasednasinferenceoptimized}. Second, the prefill and decode phases exhibit distinct compute and memory access characteristics, which exacerbate hardware underutilization. As shown in Fig.~\ref{motivation}, operations within an LLM layer generally lie in the compute-bound regime during the prefill phase, since the model processes the full input sequence. In contrast, during the decode phase with batch size 1, all operations within a layer become memory-bound, leading to severe hardware underutilization. Increasing the batch size mitigates this issue by shifting some operations toward the compute-bound regime. However, the attention layer remains memory-bound because each unique input sequence requires a separate Key-Value (KV) cache \cite{vaswani2017attention}. To tackle this hardware underutilization, several studies \cite{attacc, cent, xformer} have proposed solutions. For instance, X-Former \cite{xformer} proposes a fully on-chip CiM-based accelerator using hybrid non-volatile memory and CMOS technology, with a focus on accelerating the attention mechanism using intralayer sequence blocking. More recently, AttAcc \cite{attacc} proposes a heterogeneous system combining compute-in-DRAM (CiD) with a GPU, mapping only the attention layer to the CiD system during the decode phase. In contrast, CENT \cite{cent} employs a fully CiD based system and maps both the prefill and decode phases onto the CiD hardware. 

However, these approaches remain limited in practice. AttAcc primarily focuses on configurations with high batch sizes, which overlooks the performance bottlenecks of low-batch inference. CENT, on the other hand, only evaluates small input context lengths (e.g. 512 tokens), making it unsuitable for today's long context LLMs \cite{du2025prefillonly}. X-Former mainly focuses on optimizing the attention layer and does not consider the bottlenecks under low-batch inference. Furthermore, increasing the batch size is not always beneficial; it does not resolve the memory bottleneck in the attention layer and is often infeasible for resource constrained edge devices \cite{alizadeh2024llmflashefficientlarge}.

To address these challenges, we propose \papernamewospace, a heterogeneous Compute-in-DRAM (CiD) and Compute-in-Memory (CiM) accelerator for efficient low-batch LLM inference. \papername integrates compute units within the HBM to accelerate the decode phase. In addition, an on-chip analog CiM accelerator is co-packaged with the HBM through 2.5D integration, enabling high bandwidth acceleration of the prefill phase. Finally, vector units in the HBM logic die are used to execute non-GEMM operations. The main contributions of this work are as follows: 

\begin{itemize}

\item We propose and design a heterogeneous CiD/CiM accelerator and introduce a phase-aware mapping strategy for efficient low-batch LLM inference (Section~\ref{pa}).
\item We analyze the performance of LLMs during the prefill and decode phases on two memory centric accelerators: a fully CiD accelerator and a fully on-chip analog CiM accelerator (Section~\ref{comparison}).
\item We evaluate \papername on LLaMA-2 7B and Qwen3 8B, demonstrating up to 2.5$\times$ speedup over CENT and 18$\times$ over AttAcc (Section~\ref{perf}).


\end{itemize}

\section{Background}\label{background}
\textbf{Large Language Models:} LLMs are built on the transformer architecture \cite{vaswani2017attention}, which has emerged as the dominant backbone for modern NLP tasks \cite{qin2025largelanguagemodelsmeet}. These models scale to billions of parameters and enable a wide variety of applications ranging from conversational assistants and text summarization to code generation and multimodal reasoning \cite{github2025copilot,summarization}. The rapid growth in model size and capability, however, has been accompanied by significantly increased requirements in memory capacity, compute throughput and interconnect bandwidth. 

The inference process of an LLM can be divided into two distinct phases: prefill and decode. In the prefill phase as shown in Fig.~\ref{background} (a), the model processes the entire input sequence of length $L_{in}$, which involves executing general matrix-matrix multiplications (GEMMs) across all transformer layers. This phase is highly compute-intensive due to the large volume of operations in the input context. The performance of LLM inference in the prefill phase is commonly measured using \textit{Time-To-First-Token (TTFT)}, which captures the time required for the model to process the entire input sequence and generate the first output token. In contrast, the decode phase is autoregressive (Fig.~\ref{background} (b)); the model generates one token at a time until it produces the output sequence of length $L_{out}$. Each step of the decode phase requires reusing the cached Key-Value (KV) pairs from previous tokens and performing general matrix-vector multiplications (GEMVs). Consequently, while prefill phase is compute-bound, decode especially under batch size of  becomes memory bound, with hardware utilization dropping significantly. The performance of this phase is typically measured using \textit{Time-Per-Output-Token (TPOT)}, which measures the latency of generating each subsequent token.

\textbf{Compute in Memory Accelerators: } Analog CiM accelerators typically can implement GEMV operations using a weight stationary dataflow \cite{cimsurvey}. In this approach, the weights of a neural network are stored in the memory array in a bit-sliced format, where each memory cell can store a portion of the weight bits. The input vector is then serialized into a bit-stream and applied over multiple cycles to the wordlines of the array. For each input cycle, the resulting analog accumulation along the bit lines produces partial sums, which are converted into digital signals using analog-to-digital converters (ADCs). Outputs from multiple crossbars are subsequently combined through shift-and-add operations to reconstruct the final result. Here, the term bit-slice refers to the number of weight bits that can be stored in a memory cell, while bit-stream refers to the serialized representation of the input vector applied across cycles.

\begin{figure}[t]
\centering
\includegraphics[width=0.45\textwidth]{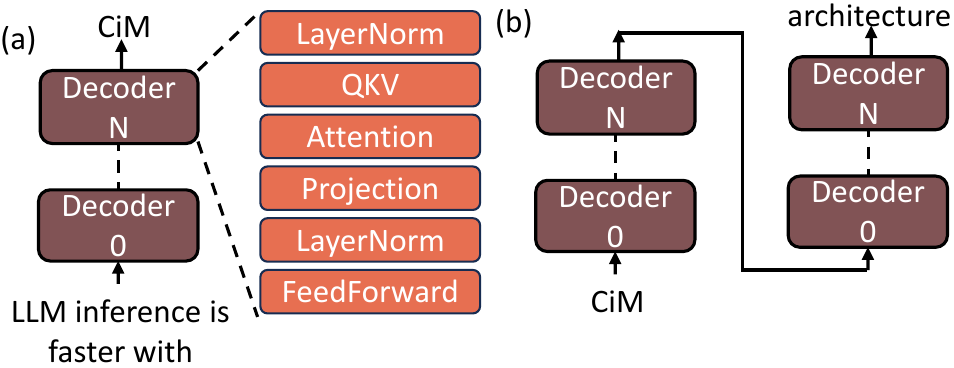}
    \caption{(a) Prefill phase of the LLM inference, where each decoder block consists of sub-operations such as LayerNorm, QKV generation, attention, projection and feedforward layers. (b) Decode phase of the LLM inference, which generates one token at a time and reuses cached Key-Value (KV) states.}

\centering
\label{background}
\vskip -0.2 in
\end{figure}

\begin{figure*}[t]
\centering
\includegraphics[width=\textwidth]{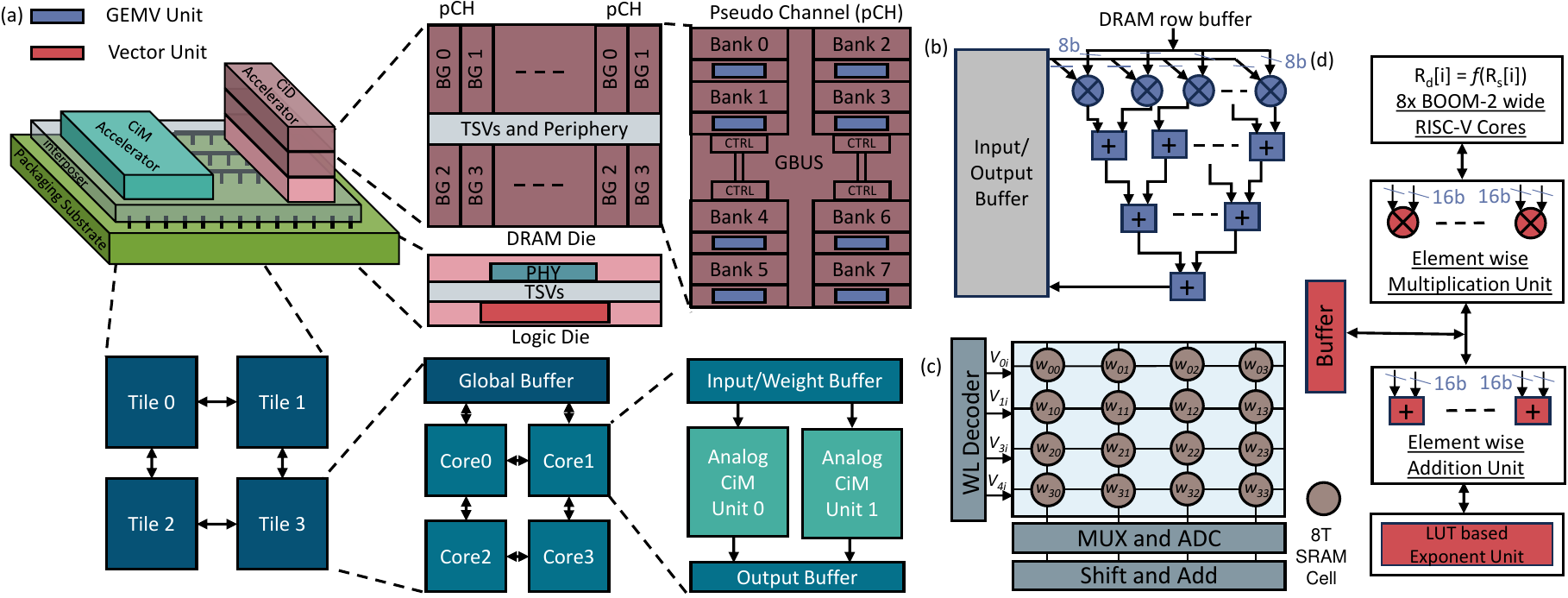}
    \caption{(a) Overview of the proposed 2.5D integrated heterogeneous accelerator architecture (\papernamewospace). The system integrates compute units within the HBM3 stack to accelerate GEMV operations, and analog compute-in-memory (CiM) accelerator co-packaged on the interposer to accelerate GEMM operations. The vector units are added to the logic die to perform non-GEMM operations. (b) Details of the GEMV units in CiD architecture. (c) Analog CiM array based on 8T SRAM cells. (d) Details of the vector units in the logic die.}


\centering
\label{arch}
\vskip -0.2 in
\end{figure*}

\section{Related Works}\label{rw}

\textbf{Fully CiD Accelerators:} Past works have explored fully CiD-based accelerators to exploit the high internal bandwidth of DRAM for transformer models \cite{transpim, cent}. TransPIM \cite{transpim} introduces an HBM based CiD architecture tailored to encoder-only models such as BERT, employing token-based data mapping to parallelize execution. CENT \cite{cent} extends this direction by designing a GPU-free system with compute eXpress link (CXL) enabled CiD devices for end-to-end LLM inference. However, these designs either restrict their scope to encoder-only models or evaluate inference scenarios with high batch sizes and short input context lengths, where prefill latency is not the critical bottleneck. In contrast, our work first analyzes the performance of LLMs on both a fully CiD accelerator and a fully on-chip analog CiM accelerator, and then demonstrates that \papername achieves superior efficiency for long-context, low-batch inference compared to CENT. 


\textbf{Heterogeneous Accelerators:} Researchers have also proposed mapping LLMs onto heterogeneous systems that combine GPUs with CiD devices \cite{attacc, neupim}. AttAcc \cite{attacc} consists of 8 A100 GPUs with HBM3 memory alongside 8 HBM-CiD devices, mapping the attention layer to CiD units during the decode phase. NeuPIM \cite{neupim} integrates a TPUv4-like architecture with CiD modules and employs dual row buffers to enable concurrent memory accesses by CiD and tensor processing unit (TPU). However, these designs primarily target high-batch inference and focus only on accelerating the attention layer. In contrast, \papername addresses low-batch LLM inference and demonstrates that non-attention layers can also become performance bottlenecks. Moreover, we compare our memory-centric heterogeneous design against a systolic-array baseline to quantify the benefits of incorporating analog CiM units for reducing prefill latency.


\section{Proposed Approach}\label{pa}
\subsection{\papernamewospace}\label{halo}

Fig.~\ref{arch} presents an overview of the proposed heterogeneous accelerator architecture, \papernamewospace, which integrates HBM based CiD and on-chip analog CiM through 2.5D integration. By co-packaging the CiM accelerator with the HBM stack on a common interposer, the design ensures high bandwidth, low-latency communication between memory and compute units. This heterogeneous integration allows \papername to efficiently support both memory-bound and compute-bound operations in LLM inference.


Within the HBM stack, compute units are embedded at the bank level to exploit fine-grained parallelism and reduce energy consumption. Placing compute units directly inside the HBM minimizes data movement across peripheral interfaces, leading to significantly lower access energy compared to off-chip computation. Each CiD-enabled bank integrates 32 8-bit multipliers. These multipliers operate in parallel. One of the inputs of the multiplier is stored in double-buffered local SRAM buffer of size 4KB (to fit 4096 8 bit inputs) this input is broadcasted to multiple bank groups and banks \cite{he2020newton}. This enables efficient execution of general matrix-vector (GEMV) operations. We also show the performance comparison of a GEMM and GEMV operation in this CiD unit in Section~\ref{comparison}. To perform the reduction operation in the GEMV operation a reduction tree is implemented within the bank itself, combining partial sums before sending it to the vector units for non-GEMM operations.

Complementing the CiD accelerator, \papername integrates an analog CiM accelerator to efficiently handle compute-bound operations. The analog CiM accelerator architecture is hierarchically organized into tiles interconnected by a 2D mesh network-on-chip (NoC). Each tile consists of multiple cores, which are themselves connected via a local 2D mesh. Each core integrates several analog CiM units that perform matrix-vector multiplications. The CiM units employ 8T SRAM based array \cite{cicc}, where one tensor of the GEMM operation is stored in a bit-sliced format and the input tensor is bit-streamed over multiple cycles. The outputs are digitized using ADCs, and shift-and-add units reconstruct the final result. This organization allows the CiM accelerator to exploit massive parallelism, making it well suited for large GEMM operations.


In addition to GEMM and GEMV acceleration, \papername incorporates vector and scalar functional units in the HBM logic die to support non-GEMM operation. Non-GEMM operations collectively account for a much smaller fraction of the overall FLOP count compared to GEMM operations, and therefore do not require massive parallelism available at the bank level. For this reason, placing these units in the logic die is sufficient to achieve low latency execution without incurring the area and energy costs of embedding them at the bank level. Vector units handle element-wise multiplications and additions required in layers such as LayerNorm and activations. Dedicated exponent units accelerate exponential functions in softmax operation, while a RISC-V BOOM core \cite{cent} is integrated to execute more general purpose arithmetic operations, including division and square root. 

\subsection{Prefill and Decode Phase Mapping}\label{mapping}

\begin{figure}[t]
\centering
\includegraphics[width=0.4\textwidth]{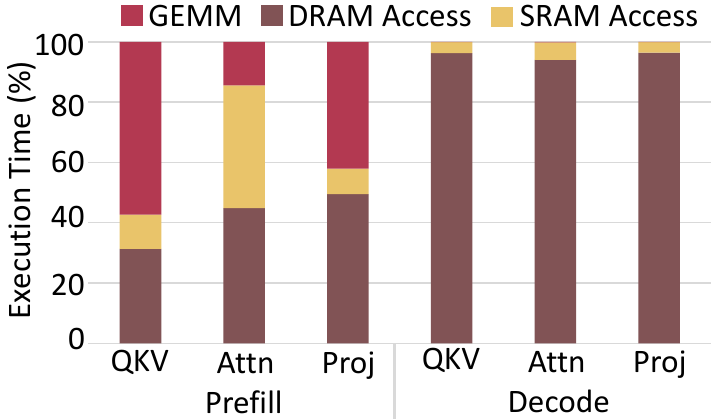}
    \caption{Execution time breakdown of different operations in the LLaMA-2 7B for prefill and decode phases with $L_{in}$=2048, $L_{out}$=128 and batch size=1.}
\centering
\label{mappingfig}
\vskip -0.2 in
\end{figure}

We profile the execution time of different operations in the LLaMA-2 7B model on the analog CiM accelerator of \papernamewospace, as shown in Fig.~\ref{mappingfig}. During the prefill phase, nearly 50\% of the execution time is consumed by GEMM operations, as the model processes multiple input tokens, pushing the workload into the compute-bound regime. In contrast, during the decode phase, where the model generates one token at a time, almost 90\% of the execution time is dominated by memory accesses to the DRAM, making the workload strongly memory-bound. 

Based on these observations, we adopt a phase-aware mapping strategy: all GEMM operations in the prefill phase are mapped to the analog CiM accelerator, while all GEMV operations in the decode phase are mapped to the CiD accelerator. Non-GEMM operations are offloaded to the logic-die vector and scalar units, as they are typically performed after results are aggregated from GEMM/GEMV operations and require minimal parallelism. In Section~\ref{comparison}, we further compare this phase-aware mapping with fully CiD and fully CiM baselines to demonstrate the performance and energy benefits of \papername. 


\section{Evaluation}\label{eval}

\subsection{Methodology}\label{method}


We estimate the latency and energy of CiD based execution using the simulator from \cite{attacc}, which we extended to support GEMM operations as well. The energy, latency and area of the 8-bit multipliers and adder trees are obtained using Cadence Genus synthesis at 65nm and then scaled to 7nm following predictive technology models \cite{scaling}. Similar to \cite{attacc}, we scale the area overhead of arithmetic units and buffers on the DRAM die to the third generation of 10nm-class (1z-nm) DRAM process \cite{park2022192}, assuming a 10$\times$ density gap between DRAM and logic processes of the same feature size. CiD compute units are added at the bank level and replicated across all the bank groups and channels. The combined area overhead of the compute units and the local SRAM buffer remains below 10\%.

\addtocounter{figure}{2}

\begin{figure*}[!b]
\centering
\includegraphics[width=\textwidth]{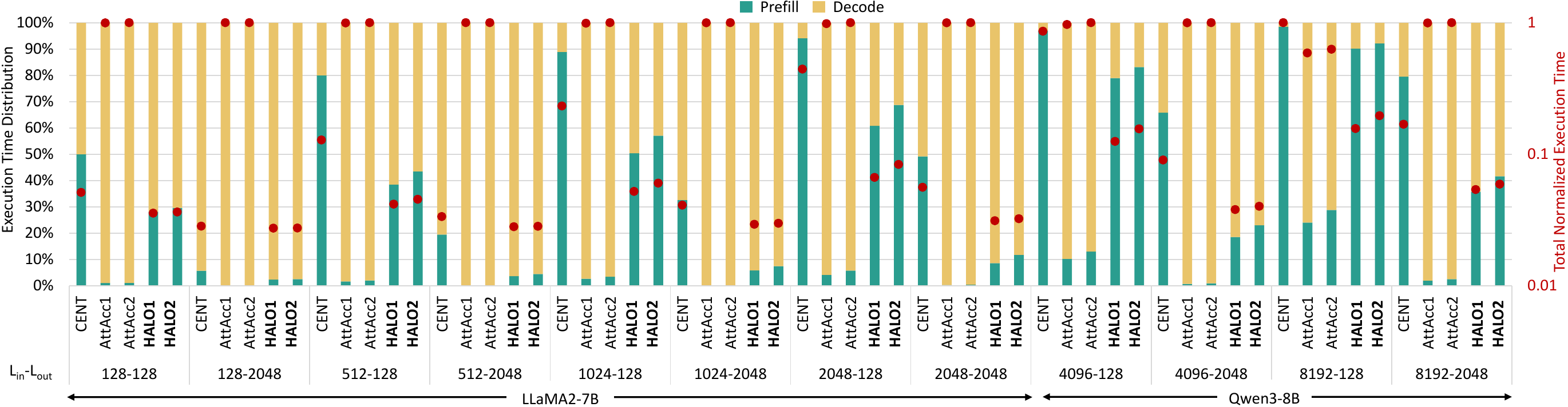}
    \caption{End-to-end execution time distribution across prefill and decode phases for LLaMA-2 7B and Qwen3 8B models. Stacked bars show the relative contribution of each phase, while red dots indicate the total normalized execution time. Normalization is performed with respect to the slowest baseline for each ($L_{in}$, $L_{out}$) configuration. Results are shown for batch size = 1, comparing our \papernamewospace1 and \papernamewospace2 mappings against prior baselines.}
    


\centering
\label{exectime}
\end{figure*}

\begin{table}[]
\centering    
\caption{\papername configuration.}
\label{spec}
\begin{tabular}{|c|c|}
\hline
\textbf{Parameter}          & \textbf{Value}                      \\
\hline\hline
HBM3               & 80 GB (5 stacks)           \\
\hline
Tile (mesh)        & 4x4                        \\
\hline
Core (mesh)        & 2x2                        \\
\hline
Global Buffer (GB) & 4 MB (2TB/s)               \\
\hline
Input Buffer (IB)  & 32 KB (4TB/s)              \\
\hline
Weight Buffer (WB) & 64 KB (4TB/s)              \\
\hline
Output Buffer (OB) & 128 KB (4TB/s)             \\
\hline
Analog CiM Unit    & 8 crossbars (128x128)      \\
\hline
ADC                & SAR, 7-bit, 48 ADC/crossbar\\
\hline
Vector Unit Width & 512 \\
\hline
\end{tabular}
\end{table}

\begin{table}[]
\centering    
\caption{Different mappings description.}
\label{mapconfig}
\begin{tabular}{|l|p{0.65\columnwidth}|}
\hline\
\textbf{Name}        & \textbf{Explanation}                                                         \\ \hline\hline
AttAcc1 \cite{attacc}      & Prefill phase on CiM (128 wordlines turned ON for 128x128 crossbar) and Attention layer during decode phase on CiD \\ \hline
AttAcc2 \cite{attacc}      & Prefill phase on CiM (64 wordlines turned ON for 128x128 crossbar) and Attention layer during decode phase on CiD \\ \hline
CENT \cite{cent}       & All the layers on CiD during prefill and decode phase               \\ \hline
\textbf{HALO1 (ours)} & Prefill on CiM accelerator (128 wordlines turned ON for 128x128 crossbar) and Decode phase on CiD accelerator (phase-aware mapping)     \\ \hline
\textbf{HALO2 (ours)} & Prefill on CiM accelerator (64 wordlines turned ON for 128x128 crossbar) and Decode phase on CiD accelerator (phase-aware mapping)    \\ \hline
\end{tabular}
\vskip -0.1 in
\end{table}

For the analog CiM accelerator in \papernamewospace, we use the analytical simulator COMET \cite{comet} to estimate latency and energy. The energy, latency and area characteristics of 8T SRAM based crossbar are derived from \cite{cicc}, while those of 7-bit SAR ADCs are taken from \cite{adc}. The architectural parameters used for \papername are summarized in Table~\ref{spec}. We compare our phase-aware mapping strategy with baseline mappings proposed in prior works including AttAcc \cite{attacc} and CENT \cite{cent}. A summary of all mapping configurations is presented in Table~\ref{mapconfig}. Due to the analog nature of computation in CiM accelerators, circuit non-idealities may degrade computation accuracy \cite{verma2019memory}. However, this degradation can be mitigated by controlling the number of wordlines activated simultaneously in the crossbar array \cite{cicc}. To explore this trade-off, we consider two configurations: \papernamewospace1 and AttAcc1, where all 128 wordlines are activated, and \papernamewospace2 and AttAcc2, where only 64 wordlines are activated. While reducing the number of active wordlines increases computation latency, it significantly mitigates the impact of circuit non-idealities on the computation accuracy \cite{cicc}. 

To understand the impact of analog CiM accelerators, we also evaluate the performance of \papername when the analog CiM crossbars are replaced with digital systolic arrays. This comparison is made at iso-area. The area, latency and energy characteristics of the systolic arrays are obtained from \cite{hisim}. All experiments are conducted on two representative LLMs: LLaMA-2 7B \cite{llama2} and Qwen3 8B \cite{yang2025qwen3}, covering both the prefill and decode phases of inference. These models were chosen to enable evaluation across a wide range of context lengths, spanning from 128 up to 10K tokens for both input and output. For all the experiments in this section, we consider a batch size of 1 and we vary the input and output context lengths. 



\subsection{Fully CiD vs Fully CiM comparison}\label{comparison}

We begin by analyzing the performance of the LLaMA-2 7B model when mapped entirely onto either the CiD or CiM accelerator in \papernamewospace. Fig.~\ref{ttft} (a) illustrates the TTFT for different input context lengths. Fully mapping the model onto the CiM accelerator yields a geometric mean speedup of 6$\times$ in TTFT compared to the fully CiD configuration. Moreover, as shown in Fig.~\ref{ttft} (b), the inference energy during the prefill phase is also lower for the CiM based execution, achieving a geometric mean reduction of 2.6$\times$ relative to the CiD mapping. This energy advantage arises because the prefill phase is compute-bound, and the CiM accelerator is more effective at local data reuse. In contrast, the CiD architecture is constrained by limited compute capability and buffer capacity, which restricts reuse and increases DRAM access energy.

Fig.~\ref{tpot}(a) presents the TPOT for the LLaMA-2 7B model across different combinations of input and output context lengths. When the model is fully mapped onto the CiD accelerator, we observe a geometric mean speedup of 39$\times$ in TPOT compared to the fully CiM configuration. This significant speedup comes from the fact that the decode phase is memory bound, and executing the operations in the CiD accelerator reduces DRAM access latency by adding compute near the DRAM banks. Additionally, decode-phase energy consumption is 3.9$\times$ lower in the CiD configuration, as data movement is minimized when computation occurs directly within the HBM. 

These results support the need for the phase-aware mapping strategy introduced in Section~\ref{mapping}, which assigns compute-bound prefill phase to CiM and memory-bound decode phase to CiD.

\addtocounter{figure}{-3}   
  
\begin{figure}[H]
\centering
\includegraphics[width=0.48\textwidth]{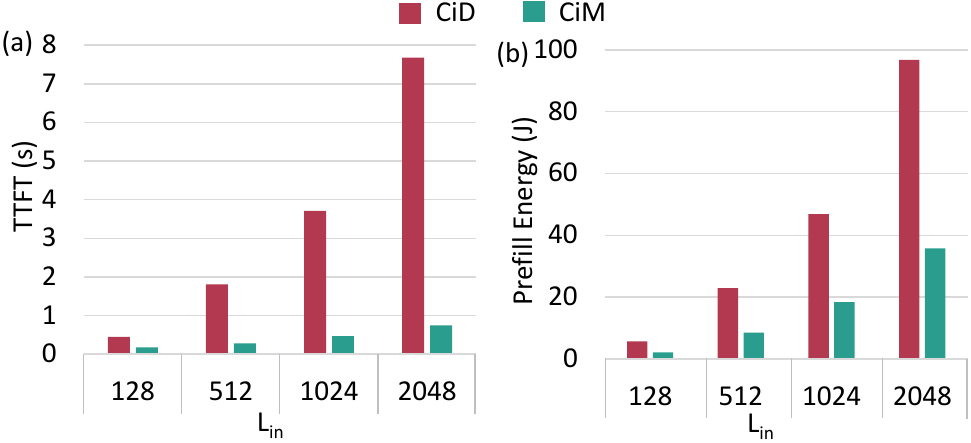}
    \caption{(a) TTFT and (b) Prefill phase energy for LLaMA-2 7B model under varying input context lengths, when mapped to fully CiD and fully CiM accelerator architecture.}
\centering
\label{ttft}
\end{figure}

\begin{figure}[H]
\centering
\includegraphics[width=0.48\textwidth]{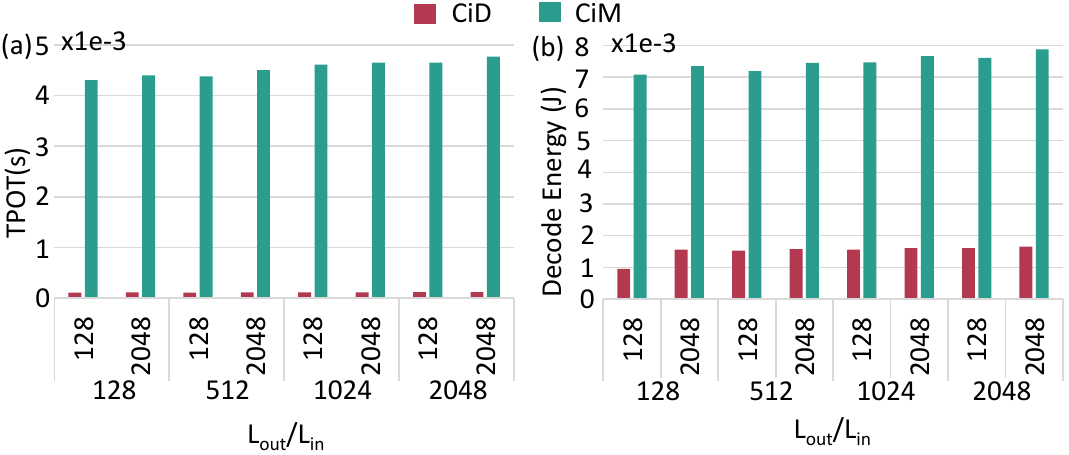}
    \caption{(a) TPOT and (b) Decode phase energy (per token) for LLaMA-2 7B model under varying input context lengths, when mapped to fully CiD and fully CiM accelerator architecture.}
\centering
\label{tpot}
\end{figure}

\subsection{Performance and Energy Comparison with Prior Works}\label{perf}



In this section, we compare the end-to-end execution time and energy of LLaMA-2 7B and Qwen3 8B models and analyze how the execution time and the energy are distributed between the prefill and decode phases. The mapping configurations used for comparison are summarized in Table~\ref{mapconfig}. Fig.~\ref{exectime} presents the execution time distribution and the total normalized execution time across various input and output context lengths. \papernamewospace1 achieves a geometric mean speedup of 6.54$\times$ in the prefill phase compared to CENT \cite{cent}. This benefit becomes even more pronounced at large input context lengths, as can be seen from Fig.~\ref{exectime}. The speedup is driven by our phase-aware mapping strategy, which assigns the compute-bound prefill phase to the CiM accelerator. In contrast, CENT maps the prefill phase to the CiD accelerator, which suffers from limited compute capability and buffer reuse. For the decode phase, \papernamewospace1 achieves similar performance to CENT, since both mappings execute the decode phase on the CiD accelerator. However, compared to AttAcc1 \cite{attacc}, \papernamewospace1 delivers a geometric mean speedup of 34$\times$. Note, AttAcc maps only the attention layer in the decode phase to CiD while the remaining operations execute on an analog CiM accelerator. In contrast, \papername executes all decode operations on CiD, significantly reducing latency. 

\addtocounter{figure}{1}
\begin{figure}[t]
\centering
\includegraphics[width=0.5\textwidth]{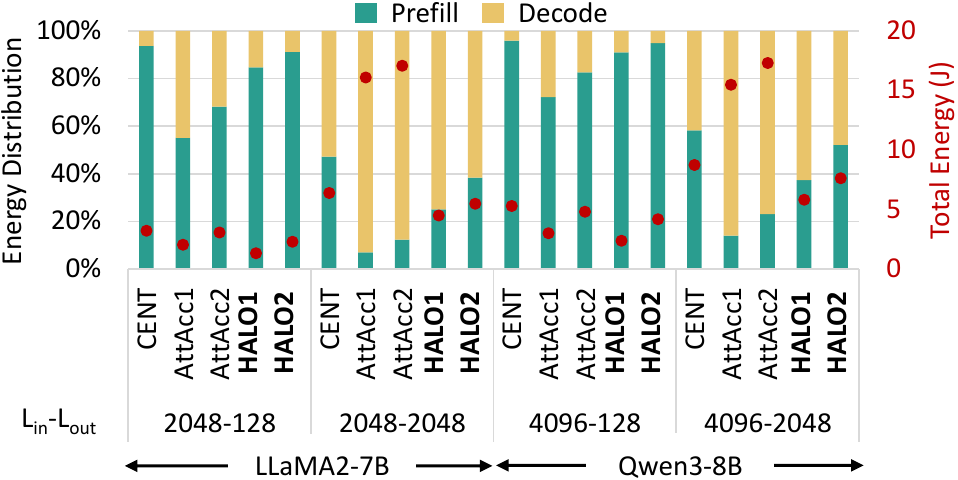}
    \caption{Total energy distribution across prefill and decode phase for LLaMA-2 7B and Qwen3 8B models. Stacked bars show the relative contribution of each phase, while red dots indicate the total energy. Results are shown for batch size = 1, comparing our \papernamewospace1 and \papernamewospace2 mappings against prior baselines.}
\centering
\label{energydist}
\end{figure}


Fig.~\ref{exectime} also shows the total execution time for both LLaMA-2 7B and Qwen3 8B across a range of ($L_{in}$, $L_{out}$) configurations. For small input context lengths, \papernamewospace1 performs similarly to CENT because the prefill phase contributes less to the overall execution time. However, for large input context lengths(512-8192), \papername consistently outperforms CENT due to its efficient prefill mapping. Interestingly, AttAcc outperforms CENT at extreme configurations such as very high input context length (e.g. 4096, 8182) and very low output context length (e.g. 128). Overall, \papernamewospace1 achieves an 18$\times$ geometric mean speedup over AttAcc1 and a 2.4$\times$ speedup over CENT in end-to-end execution time.


\begin{wrapfigure}{r}{0.25\textwidth}  
    \centering
    \vspace{-8pt} 
    \includegraphics[width=0.24\textwidth]{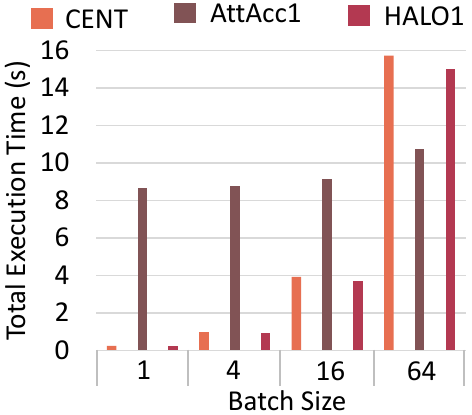}
    \caption{Execution time comparison of the LLaMA-2 7B model across different baselines as batch size varies, with $L_{in}=128$ and $L_{out}=2048$.}
    \label{bs}
    \vspace{-5pt}
\end{wrapfigure}

Finally, we also evaluate a CiM configuration, \papernamewospace 2, where only 64 of 128 wordlines in the crossbar are activated. This design reduces the impact of circuit non-idealities \cite{cicc} but increases compute time. We observe only a 10\% geometric mean slowdown relative to \papernamewospace1. The small degradation is due to the longer compute latency being amortized by improved overlap with parent memory (GB) fills into the child buffers (IB, WB, OB) \cite{comet}, maintaining efficient pipeline utilization.



Fig.~\ref{bs} shows the end-to-end execution time of the LLaMA-2 7B model on \papername compared to prior mappings optimized for higher batch sizes. At lower batch sizes ($<$ 64), both \papernamewospace1 and CENT achieve lower latency due to the decode phase being memory-bound. As batch size increases (e.g. 64), AttAcc1 \cite{attacc} becomes more effective, since non-attention operations become compute-bound and benefit from CiM acceleration.

Fig.~\ref{energydist} shows the total energy distribution and overall energy consumption for different input and output context lengths. We observe that \papernamewospace1 achieves a 2$\times$ geometric mean reduction in energy compared to AttAcc1, primarily due to its lower decode energy, as illustrated in the energy distribution bar plot. When compared to CENT, \papernamewospace1 delivers a 1.8$\times$ geometric mean reduction in energy, which stems from improved data reuse in the prefill phase on the analog CiM accelerator in contrast to the CiD accelerator. Finally, \papernamewospace2 exhibits higher energy consumption than \papernamewospace1 and is comparable to CENT, this is due to the double ADC accesses incurred when only half of the wordlines are activated in the analog CiM accelerator.

\subsection{Performance Comparison with Digital Accelerator}

\begin{figure}[t]
\centering
\includegraphics[width=0.5\textwidth]{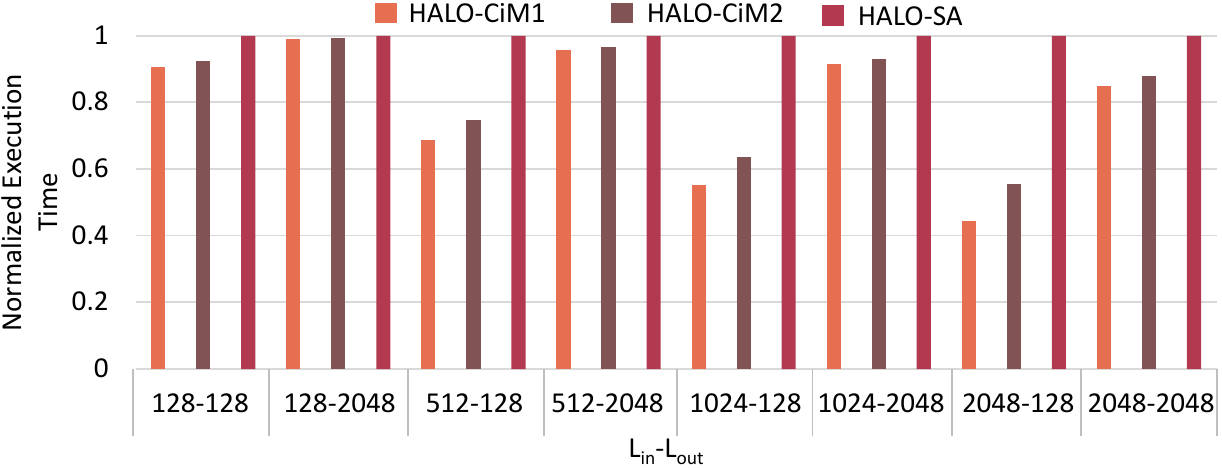}
    \caption{Normalized execution time comparison of \papername with analog CiM crossbars (\papernamewospace-CiM1,2) vs digital systolic arrays (\papernamewospace-SA) for the LLaMA-2 7B model across various ($L_{in}$, $L_{out}$) configurations (batch size=1).}
\centering
\label{sacomp}
\end{figure}

To evaluate the impact of our memory-centric design, we replace the analog CiM units in \papername with digital systolic arrays, resulting in a NeuPIM architecture \cite{neupim}. Specifically, we use two 128x128 systolic arrays per core in Fig.~\ref{arch}, supporting 8b x 8b MAC operations, while maintaining iso-area with the \papernamewospace-CiM configuration. Fig.~\ref{sacomp} presents the normalized execution time for the LLaMA-2 7B model across varying input and output context lengths. We observe a geometric mean speedup of 1.3$\times$ and 1.2$\times$ for \papernamewospace-CiM1 and \papernamewospace-CiM2, respectively, compared to systolic array based design (\papernamewospace-SA). These results demonstrate the performance advantage of analog CiM in compute-bound phases, highlighting the effectiveness of memory-centric integration in reducing execution time.

\section{Conclusion}\label{conclusion}


In this work, we characterize the performance of LLMs during prefill and decode phases under low-batch inference settings. We observe that the prefill phase is compute-bound, with multiple GEMM operations dominating execution time. In contrast, during the decode phase, not only the attention operations but also others such as QKV generation, projection layers, and feed-forward layers become memory-bound due to the lack of batch-level parallelism. To address these phase specific bottlenecks, we propose \papername, a heterogeneous CiD/CiM accelerator for efficient low-batch LLM inference. \papername integrates compute units within HBM stacks to accelerate GEMV operations, and co-packages an on-chip analog CiM accelerator using 2.5D integration to enable high bandwidth GEMM execution. We further introduce phase-aware mapping strategy that adapts the LLM execution to the characteristics of each phase. Our experimental results on LLaMA-2 7B and Qwen3 8B models mapped to \papername achieves 18$\times$ and 2.5$\times$ speedup over other state of the art accelerators such AttAcc and CENT.

\section*{Acknowledgment}

This work was supported in part by the Center for the Co-Design of Cognitive Systems (COCOSYS), a DARPA-sponsored JUMP center of Semiconductor Research Corporation (SRC), in part by the National Science Foundation, in part by Intel Corporation and in part by the Department of Energy. The authors would also like to thank Pragnya Nalla (University of Minnesota) for providing the area numbers for the systolic array.

\bibliographystyle{IEEEtranS}
\bibliography{refs}

\end{document}